\documentclass[conference]{IEEEtran}

\usepackage{cite}
\usepackage{graphicx}
\usepackage{url}
\usepackage{amsmath,amssymb,amsfonts}
\usepackage{enumitem}
\usepackage[table]{xcolor}
\usepackage{booktabs}
\usepackage{multirow}
\usepackage{balance}
\usepackage{newtxtext}

\newcommand{\sievefl}{\textsc{SieveFL}}
\newcommand{\vone}{$\text{V}_1$}
\newcommand{\vzero}{$\text{V}_0$}
\newcommand{\vtwo}{$\text{V}_2$}

\newcommand{\eg}{\textit{e.g.,}}

\title{
SieveFL: Hierarchical Runtime-Aware Pruning for Scalable LLM-Based Fault Localization
}
\author{
  Mahdi Farzandway, Fatemeh Ghassemi \\
  University of Tehran \\
  \texttt{\{mahdifarzandway, fghassemi\}@ut.ac.ir} \\
}

\begin{document}

\maketitle

\begin{abstract} 
Automated fault localization requires connecting an observed test failure to the responsible method across thousands of candidates—a task that purely statistical approaches handle with limited precision and that LLMs cannot yet handle at full project scale due to prohibitive token cost and signal dilution. We present SieveFL, a five-stage hierarchical framework that resolves this tension through aggressive pre-LLM filtering. SieveFL converts a failing test into a natural-language failure description, uses dense vector retrieval to narrow the search to a small set of suspicious files, and then eliminates any method not executed during the failing test via JaCoCo runtime traces. Only the surviving candidates are passed to the LLM, which screens each method individually and re-ranks the confirmed suspects in a single comparative pass.
We evaluate SieveFL on 395 bugs from Defects4J v1.2.0 using a mid-sized, openly available MoE model deployed on a commodity workstation (32 GB RAM, 8 GB GPU) via Ollama—no frontier APIs or datacenter hardware required. Treating 12 incomplete runs as failures, SieveFL achieves Top-1 accuracy of 41.8\% (165/395 bugs) and an MRR of 0.469, outperforming the strongest prior agent-based baseline (AgentFL) by 2.1 pp in Top-1. Runtime pruning removes 79\% of candidate methods and reduces input token consumption by 49\%, while simultaneously improving ranking quality: Top-1 is preserved exactly and Top-3 through Top-10 improve by up to 2.4 pp. These results demonstrate that, with the right filtering architecture, capable fault localization does not require proprietary frontier models. 
\end{abstract} 

\section{Introduction}
\label{sec:introduction}

Software debugging consumes a disproportionate share of developer time, with fault localization often cited as one of the most demanding phases of the entire development lifecycle~\cite{wong2016survey,kochhar2016practitioners}. The core difficulty is not computational but conceptual: connecting an observed failure to the specific method responsible requires understanding what the code was supposed to do, not just what it did. Spectrum-Based Fault Localization (SBFL) approaches such as Ochiai~\cite{abreu2007accuracy} approximate this by correlating execution patterns with test outcomes, while Information Retrieval approaches~\cite{zhou2012should,saha2013improving,li2020irbfl} treat the problem as document matching between bug reports and source indices. Both families are efficient, but neither can bridge the semantic gap between a raw execution anomaly and the architectural intent behind the failing code.

Large Language Models have changed what is possible here. Recent systems including AutoFL~\cite{kang2024quantitative}, AgentFL~\cite{qin2024agentfl}, and FaR-Loc~\cite{shi2025enhancing} demonstrate that LLMs can reason causally about code in ways that statistical or retrieval tools cannot, essentially simulating the reasoning process of an experienced developer reading a stack trace. The practical obstacle is scale. A real-world project may expose thousands of candidate methods to the model, and the relevant signal quickly drowns in noise. Irrelevant methods that happen to share vocabulary with the failure context accumulate false positives, while the token cost of evaluating each candidate individually becomes prohibitive~\cite{shi2025enhancing}. We refer to this tension between coverage and precision as the \emph{Scale-Precision Dilemma}: the larger the codebase, the less effective LLM-based reasoning becomes if applied without prior filtering.

We present \textsc{SieveFL}, a five-stage hierarchical framework designed to resolve this dilemma through progressive pruning. The key insight is that the LLM should be the last filter applied, not the first. \textsc{SieveFL} constructs a cascade in which each stage dramatically narrows the candidate set before passing it to the next, more expensive, stage. It begins by converting raw test failures and stack traces into a natural-language failure description. Semantic retrieval then reduces the search to a small set of suspicious files. A JaCoCo runtime trace eliminates any method that was never executed during the failing test. Only after these three filtering steps does the LLM become involved: it screens each surviving method individually for plausibility, then ranks the confirmed suspects in a single comparative pass. Critically, the entire framework runs on a mid-sized, openly available MoE model (Nemotron-3-Nano-30B-A3B) via Ollama on a commodity workstation, without relying on frontier APIs or datacenter-class infrastructure, which demonstrates that the accuracy gains come from the architecture, not from model scale.

\textbf{Contributions.} Our primary contributions are:
\begin{itemize}[leftmargin=*]
    \item We propose a \textbf{five-stage hierarchical pruning architecture} that progressively reduces the candidate search space from the full codebase to a small ranked list of suspicious methods. Runtime-aware pruning via JaCoCo reduces the candidate set by 79\% and LLM input token consumption by approximately 49\%, while simultaneously \emph{improving} localization quality: Top-1 accuracy is preserved exactly ($\Delta = 0.0$~pp), Top-3, Top-5, and Top-10 improve by up to 2.4~pp, and MRR increases from 0.462 to 0.469---all on the same 383-bug paired evaluation set.    
    \item We introduce a \textbf{per-method LLM screening protocol} in which each surviving candidate is evaluated in an isolated query, enabling focused causal reasoning free from inter-method interference and position bias. The binary verdicts and justifications produced at this stage serve as structured input to the final re-ranking step.
    \item We demonstrate that \textbf{state-of-the-art fault localization does not require frontier models}. Running entirely on Nemotron-3-Nano-30B-A3B, a mid-sized openly available MoE model deployable on a commodity workstation with 32\,GB of system memory and an 8\,GB consumer GPU via Ollama, \sievefl{} achieves Top-1 accuracy of 41.8\% (165/395~bugs, treating incomplete runs as failures) on the full benchmark, and MRR of 0.469 on the 383~completed bugs of Defects4J v1.2.0~\cite{just2014defects4j}, outperforming the strongest prior agent-based baseline (AgentFL) by 2.1~pp in Top-1 when both systems are evaluated over the full 395-bug benchmark.
    
\end{itemize}

\section{Related Work}
\label{sec:related_work}

\subsection{Spectrum-Based and Learning-Based Fault Localization}

Fault localization has evolved from statistical heuristics to data-driven models. Traditional SBFL metrics such as Ochiai~\cite{abreu2007accuracy} and D*~\cite{wong2013dstar} rank methods by correlating their execution coverage with test pass/fail outcomes. Although computationally efficient, SBFL techniques are sensitive to test suite quality and are known to struggle on large-scale projects with complex execution profiles~\cite{heiden2019evaluation}. Learning-based FL methods such as \textit{DeepFL}~\cite{li2019deepfl} and \textit{GRACE}~\cite{lou2021boosting} address this by integrating multiple signals---spectrum, code metrics, and syntactic features---through deep learning. \textit{GRACE} in particular uses graph-based representations to capture structural relationships between program elements. While these models achieve strong results on standard benchmarks, they require substantial labeled training data and may generalize poorly to unseen software systems, as their learned representations are inherently tied to the training distribution. A complementary challenge surfaces in multi-fault settings, where \textit{SA-BCL}~\cite{du2026augmenting} identifies two compounding deficiencies in program spectra: an imbalance of defect knowledge caused by the relative scarcity of failing test cases, and the presence of \emph{characterizing noise} introduced when mutual fault interference causes some failing executions to exhibit coverage patterns indistinguishable from passing ones. SA-BCL addresses both issues by first detecting borderline failing samples and augmenting them via random oversampling, then applying confident learning~\cite{northcutt2021confident} to identify and remove noisy spectrum entries before iterative spectrum reduction. On both the synthetic TCM and the real-world Defects4J multi-fault benchmarks, SA-BCL outperforms the state-of-the-art FLITSR by up to 37.5\% in Average Wasted Effort and 29.2\% in precision---results that confirm spectrum quality as a critical bottleneck in coverage-based localization and motivate augmenting static coverage signals with cleaner execution evidence. \textsc{SieveFL} departs from the purely spectrum-based paradigm by relying on zero-shot LLM reasoning and runtime traces, neither of which requires project-specific training. However, this training-free design comes with a trade-off. Learning-based techniques such as GRACE~\cite{lou2021boosting} and DeepFL~\cite{li2019deepfl} benefit from a leave-one-out training strategy that exposes the model to the structural and coverage patterns of the target project before evaluation. This in-project knowledge gives them a systematic advantage on projects with recurring bug patterns or structurally similar fault sites---an advantage that zero-shot LLM reasoning cannot replicate without prior exposure to the codebase. This distinction is important when interpreting the benchmark results in Section~\ref{sec:rq1}, where GRACE outperforms \sievefl{} on aggregate Top-1 accuracy despite \sievefl{} achieving superior results on projects with discriminative runtime traces.

\subsection{IR-Based Fault Localization}

IR-based FL approaches treat the bug report as a natural-language query and the source code as a document corpus~\cite{zhou2012should,saha2013improving,li2020irbfl}. BM25-based techniques~\cite{robertson2009probabilistic} retrieve files or methods whose token distributions are most similar to the bug report. While straightforward and fast, these approaches are limited by vocabulary mismatch: they cannot bridge the gap between the natural-language description of a failure and the often terse, identifier-heavy style of source code. \textsc{SieveFL}'s Stage~2 addresses this limitation by using a shared embedding space that maps both natural language and source code into comparable semantic representations, while Stages~3--4 augment the retrieval signal with dynamic execution evidence that purely textual methods cannot access.

\subsection{LLMs and RAG in Software Engineering}

The integration of LLMs has redefined automated program repair (APR) and fault localization~\cite{xia2024automated,yang2024large}. \textit{AutoFL}~\cite{kang2024quantitative} pioneered the use of LLMs to generate explainable fault evidence via function call tools, demonstrating that LLMs can perform meaningful causal reasoning about code when supplied with sufficient context. \textit{FuseFL}~\cite{widyasari2024demystifying} extends this direction by combining SBFL suspiciousness rankings, test execution outcomes, and natural-language code descriptions into a single prompt, enabling LLMs to produce step-by-step explanations of why a specific line is faulty; the approach substantially outperforms pure SBFL baselines and shows that richer contextual signals sharpen LLM localization judgments. However, the limited context window of LLMs necessitates Retrieval-Augmented Generation (RAG)~\cite{gao2023retrieval} to pre-select a manageable candidate set before LLM invocation. \textit{FaR-Loc}~\cite{shi2025enhancing} introduced functionality-aware retrieval that leverages pre-trained code embeddings to bridge the semantic gap between failure descriptions and code implementations.

A complementary direction structures LLM output rather than leaving it as free-form text. \textit{SemLoc}~\cite{yang2026semloc} converts LLM-inferred semantic properties into a closed intermediate representation of typed, executable constraints anchored to specific program locations. Executing these constraints across passing and failing tests yields a \emph{semantic violation spectrum} analogous to coverage-based SBFL, and a counterfactual verification step further distinguishes root-cause violations from cascading downstream effects. SemLoc is particularly effective on semantic bugs where identical execution paths differ only in whether a numeric invariant holds---a class of faults that syntactic spectra cannot distinguish. In the educational setting, \textit{FLAME}~\cite{liu2025explainable} demonstrates that prompting LLMs to \emph{annotate} faulty lines directly inside the source listing---rather than predicting bare line numbers---substantially improves localization accuracy and produces human-readable explanations suitable for student feedback, further illustrating that how the LLM interacts with code structure has a large effect on precision.

An important empirical constraint on context design comes from Sepidband et al.~\cite{sepidband2026role}, who conduct a large-scale factorial study of fault localization context granularity across 61 configurations on SWE-bench Verified. Their results establish three findings directly relevant to \textsc{SieveFL}: file-level context provides the dominant performance gain (a $15$--$17\times$ improvement over a no-file baseline), expanding line-level context \emph{frequently degrades} repair performance by introducing noise that dilutes the localization signal, and LLM-based file retrieval outperforms structural heuristics while incurring lower token cost. These findings empirically justify \textsc{SieveFL}'s design decision to invest in aggressive hierarchical pruning at the file and method levels before any LLM call is made, reserving the LLM for the small, high-signal survivor set.

\textsc{SieveFL} builds upon these insights but adds two critical layers absent in prior RAG-based FL work: a runtime pruning stage that intersects the semantic candidate set with dynamic execution traces, and a per-method isolated screening step that enables focused causal reasoning before comparative re-ranking. The importance of controlling retrieval noise in RAG systems has been highlighted independently~\cite{cuconasu2024power}; our results confirm this finding in the FL domain, where methods removed by runtime pruning are shown to be predominantly false-positive candidates.

\subsection{Agentic and Multi-Agent Frameworks for Debugging}

Recent work emphasizes agentic architectures to mimic the iterative reasoning of human debuggers. \textit{AgentFL}~\cite{qin2024agentfl} decomposes FL into a three-stage agentic process of test comprehension, codebase navigation, and fault confirmation, operating at project scale without requiring coverage instrumentation. Rafi et al.~\cite{rafi2024multi} propose a multi-agent system that combines graph-based code retrieval with a reflexion mechanism to refine fault hypotheses across multiple reasoning rounds. \textit{MemFL}~\cite{yeo2025improving} takes a different strategy for incorporating project-specific knowledge: rather than navigating the repository at inference time, it precomputes a two-component external memory---static summaries of the project and its constituent classes, and dynamically refined debugging guidance distilled from prior localization attempts on a small training set---which is prepended to a lightweight three-step pipeline of bug review generation, code condensation, and fault confirmation. On Defects4J, MemFL achieves substantially higher Top-1 accuracy than AutoFL and SoapFL at a fraction of their cost and runtime, with especially pronounced gains on complex projects such as Closure where generic LLM reasoning degrades. A separate thread of agentic debugging research equips LLMs with \emph{interactive dynamic analysis} rather than pre-computed summaries or multi-round reflexion over static context. \textit{InspectCoder}~\cite{wang2026inspectcoder} is the first agentic repair system to give an LLM agent direct, programmatic control over a live Python debugger. Its dual-agent architecture pairs a \textit{Program Inspector}---which strategically places breakpoints, inspects runtime variable states, and injects perturbation logic within a stateful debugging session---with a \textit{Patch Coder} that synthesizes fixes grounded in the inspector's root-cause report. Crucially, rather than passively consuming pre-collected execution logs as in log-augmented approaches such as LDB~\cite{zhong2024debug}, InspectCoder \emph{adaptively} queries runtime state in response to intermediate findings, receiving immediate process-reward signals that guide multi-step hypothesis refinement without committing to irreversible code changes. A specialized middleware, InspectWare, models the debugger as a finite-state machine and shields the LLM from low-level protocol noise, providing structured, context-aware feedback at each reasoning step. Evaluated on BigCodeBench-R and LiveCodeBench-R, InspectCoder achieves 5.10\%--60.37\% relative improvements in repair accuracy over the strongest baselines and resolves $1.67\times$--$2.24\times$ more bugs per hour---evidence that execution-grounded, hypothesis-driven exploration substantially outperforms both static reasoning and passive log collection. These results echo a principle central to \textsc{SieveFL}: dynamic execution evidence, whether gathered through interactive debugger control or test-trace intersection, greatly sharpens the signal available to the LLM and reduces the cost of each reasoning step. While these approaches demonstrate the value of structured reasoning and project-specific context, they tend to invoke the LLM over large, unfiltered candidate sets, making them vulnerable to the Scale-Precision Dilemma. Furthermore, sequential multi-agent pipelines can suffer from compounding reasoning errors across turns~\cite{shinn2023reflexion}. \textsc{SieveFL} takes a different stance: rather than relying on multi-round agent interaction or pre-baked project summaries to compensate for a noisy candidate pool, it invests in pre-LLM filtering so that each LLM call---whether the per-method screening in Stage~4 or the comparative re-ranking in Stage~5---operates on a small, high-signal candidate set where a single focused query suffices.

\section{Proposed Method}
\label{sec:method}

Fault localization---the task of identifying the source code elements responsible for an observed failure---has been studied extensively across a wide range of automated techniques~\cite{wong2016survey}. We present \sievefl{}, a five-stage hierarchical framework for precise method-level fault localization. The central design philosophy is \emph{progressive pruning}: rather than exposing all methods of a large codebase to expensive analysis at once, \sievefl{} constructs a cascade of increasingly focused filters. Each stage dramatically reduces the candidate set passed to the next, so that the most costly reasoning—individual LLM interrogation of each method—is applied only to a small, high-quality pool of suspects.

\begin{figure*}[ht]
    \centering
    \includegraphics[width=0.9\textwidth]{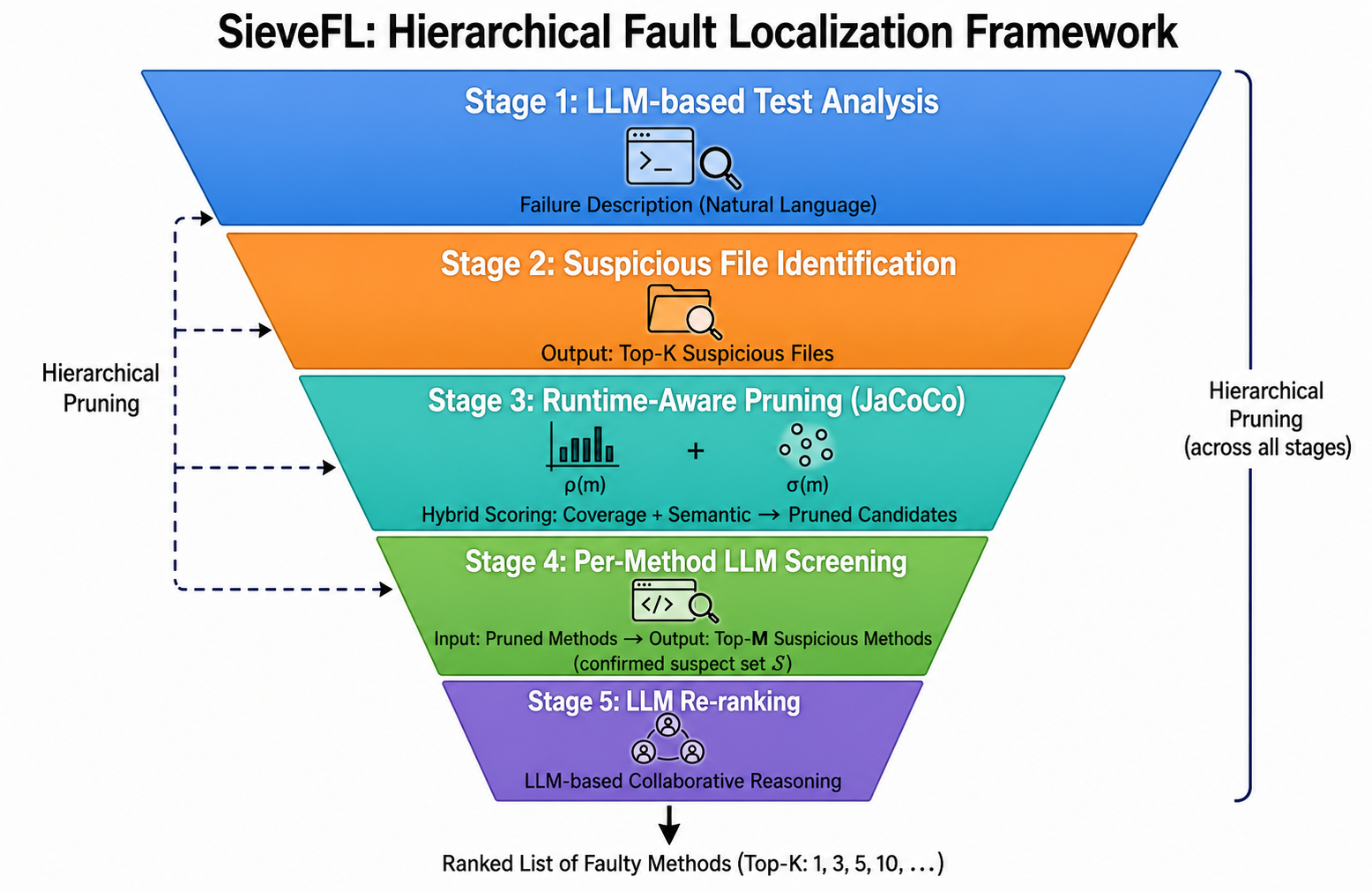}
    \caption{
        Overview of the \sievefl{} framework. The five-stage pipeline hierarchically prunes the search space from the full codebase to a ranked list of suspicious methods. Stages~1--3 perform progressively fine-grained filtering using LLM-based failure analysis, file-level retrieval, and runtime-aware pruning that combines JaCoCo coverage signals with semantic similarity. Stage~4 identifies top suspicious methods from the pruned candidates, and Stage~5 performs LLM-based collaborative re-ranking to produce the final ranked list.
    }
    \label{fig:sievefl_overview}
\end{figure*}

Figure~\ref{fig:sievefl_overview} illustrates the full pipeline,
which comprises:
(1)~\textit{LLM-based Test Analysis},
(2)~\textit{Suspicious File Identification},
(3)~\textit{Runtime-Aware Candidate Pruning},
(4)~\textit{Per-Method LLM Screening}, and
(5)~\textit{LLM-Based Re-ranking}.
Stages~1--3 convert raw failure artifacts into a semantically rich query, narrow the search space to a small set of candidate files, and then remove methods that were not exercised during the failing test. Stage~4 individually interrogates each surviving method to determine whether it is plausibly responsible for the failure. Stage~5 collects the confirmed suspicious methods and asks the LLM to rank them from most to least likely to contain the fault.

\subsection{Stage 1: LLM-based Test Analysis}
\label{sec:stage1}

The primary goal of Stage~1 is to bridge the semantic gap between a raw test failure and a natural-language description that can drive all subsequent retrieval and reasoning steps. Given the failing test function(s) and their associated error output, we prompt a large language model to perform structured reasoning over the inputs and produce a comprehensive \emph{failure description} in prose.

The failure description must capture three complementary aspects of the bug. First, it articulates the \emph{expected behavior}---the contract that the failing test asserts about the system under test. Second, it explains the \emph{observed error}---the discrepancy between the expected and actual outcomes, grounded in the specific exception type, stack trace, or assertion message. Third, and most consequentially for all downstream steps, it produces a \emph{search-oriented summary} that names the likely faulty functionality in terms that align with how a developer would describe it in code comments and method signatures.

Because this description serves as the shared query across every subsequent stage, we design the prompt (Figure~\ref{fig:prompt1}) to discourage shallow paraphrasing of the error message and instead elicit reasoning about \emph{why} the failure occurred at the semantic level.

\begin{figure}[ht]
    \centering
    \fbox{\parbox{0.44\textwidth}{%
    \small
    \textbf{System:}~You are an expert software debugger with deep knowledge of Java and common fault patterns.\\[4pt]
    \textbf{User:}~Analyze the following failing test case:
    \texttt{\{test\_code\}}.\\[4pt]
    Your response must address three points:\\
    1.~Explain the \textbf{expected behavior} of the tested functionality.\\
    2.~Explain the \textbf{actual observed failure}, including the error type and any relevant context from the stack trace.\\
    3.~Generate a concise \textbf{search query} (2--5 sentences) describing the likely faulty functionality in natural language, as if you were searching a codebase for the methods responsible for this behavior. 
    }}
    \caption{Prompt used in Stage~1 (LLM-based Test Analysis). The three-point structure elicits semantic reasoning beyond surface-level error paraphrasing.}
    \label{fig:prompt1}
\end{figure}

\subsection{Stage 2: Suspicious File Identification}
\label{sec:stage2}

Stage~2 reduces the full codebase to a small set of candidate files that are semantically aligned with the failure description produced in Stage~1. This design follows the Retrieval-Augmented Generation (RAG) paradigm~\cite{gao2023retrieval}, adapted here to treat individual methods as the retrieval corpus and the failure description as the query. This coarse-grained filtering serves two purposes. 
First, it removes the vast majority of source files irrelevant to the fault, preventing the downstream stages from being overwhelmed by unrelated candidates---a known failure mode in retrieval-augmented systems where noisy candidates degrade final output quality~\cite{cuconasu2024power}.
Second, by indexing at the method level, each document in the corpus represents a cohesive unit of behavior, which produces more semantically precise similarity scores than whole-file or fixed-size chunk representations.

\textbf{Method-level indexing.}
Rather than chunking source files arbitrarily, we parse every source file in the project and extract each method together with its associated Javadoc and inline comments as a single, self-contained document. A file containing $n$ methods therefore contributes exactly $n$ documents to the index, each carrying its parent file path as metadata. All documents are embedded using
\texttt{sentence-transformers/all-MiniLM-L6-v2}~\cite{reimers2019sentence}, a lightweight model that maps both natural language and source code identifiers, comments, and string literals into a shared semantic space. The resulting embeddings are stored in a FAISS flat
index~\cite{douze2025faiss}, and similarity is measured by cosine distance~\cite{salton1975vector}.

\textbf{File-level aggregation.}
At query time, the failure description from Stage~1 is embedded and the top-$k_f$ most similar method documents are retrieved. The \emph{suspicious file set} $\mathcal{F}^*$ is then formed by taking the union of the parent file paths of these retrieved methods---that is, retrieval is performed at method granularity, but the output is a set of files. This two-level design is deliberate: method-level matching yields precise semantic alignment, while file-level aggregation ensures that the downstream stages receive complete files for coverage analysis and exhaustive method enumeration. We set $k_f = 10$, which consistently recovers the ground-truth file in over 90\% of bugs while keeping $|\mathcal{F}^*|$ small enough for efficient downstream processing.

\subsection{Stage 3: Runtime-Aware Candidate Pruning}
\label{sec:stage3}

Semantic retrieval alone cannot distinguish between methods that are \emph{described} similarly to the failure and methods that were actually \emph{executed} during it. Stage~3 closes this gap by introducing a dynamic runtime signal: branch-level coverage traces collected by JaCoCo~\cite{jacoco}, the industry-standard Java bytecode instrumentation framework.

\textbf{Coverage collection.}
For each bug, we instrument the subject program with JaCoCo and execute only the identified failing test method(s)---not the full regression suite---so the resulting trace is specific to the fault-triggering execution path rather than an aggregate of all test activity. JaCoCo exports results in its standard XML report format
(\texttt{coverage.xml}).

\textbf{Hybrid scoring.}
Within each suspicious file $f \in \mathcal{F}^*$, we extract the method set $\mathcal{M}(f)$ by parsing the JaCoCo report. Combining multiple complementary signals has been shown to consistently outperform any single signal in isolation for fault localization~\cite{xuan2014learning}. Following this principle, for each method $m \in \mathcal{M}(f)$, we compute a hybrid score combining the runtime coverage signal with the semantic similarity from Stage~2:
\begin{equation}
  s(m) \;=\; w_{\text{cov}}\,\cdot\,\rho(m) \;+\; w_{\text{sem}}\,\cdot\,\sigma(m),
  \label{eq:hybrid}
\end{equation}
where $\rho(m) \in [0,1]$ is the branch-coverage ratio of method $m$ (fraction of its branches executed during the failing test), $\sigma(m) \in [0,1]$ is the cosine similarity between the method's embedding and the failure description from Stage~1, and $w_{\text{cov}}$, $w_{\text{sem}} \geq 0$ with $w_{\text{cov}} + w_{\text{sem}} = 1$. Our ablation study (Section~\ref{sec:rq3}) shows that retaining the semantic component is essential for both ranking precision and for protecting ground-truth methods from being pruned.

\textbf{Pruning decision.}
Any method whose score $s(m)$ falls below a threshold $\tau$ is removed before any LLM call is made. For methods affected by \emph{method-signature overload}---where JaCoCo's line-level map cannot unambiguously resolve which overloaded variant was invoked---we conservatively retain all overloaded variants. When no \texttt{coverage.xml} is available (\eg{} due to build incompatibilities), Stage~3 degrades gracefully by falling back to the \vzero{} configuration, preserving correctness at the expense of cost savings.

The output of Stage~3 is a pruned candidate set $\hat{\mathcal{M}} \subseteq \bigcup_{f \in \mathcal{F}^*} \mathcal{M}(f)$. As reported in Sections~\ref{sec:rq2} and~\ref{sec:rq3}, this step retains on average only 21\% of the methods present before pruning (reduction ratio $= 0.79$) while preserving the ground-truth method in over 91\% of bugs. This dramatic reduction is what makes the per-method LLM interrogation in Stage~4 computationally feasible.

\subsection{Stage 4: Per-Method LLM Screening}
\label{sec:stage4}

Stage~4 is the core reasoning step of \sievefl{}. Having reduced the candidate set to a manageable size in Stage~3, we now interrogate each surviving method individually using the LLM. Recent work has demonstrated that LLMs can perform meaningful causal reasoning about code faults when given sufficient and focused context~\cite{kang2024quantitative,yang2024large}. Rather than asking the model to compare all candidates simultaneously---which would dilute its attention and introduce position bias---we present each method in a dedicated query and ask a focused binary question: \emph{is this method plausibly responsible for the observed failure?}

For each method $m \in \hat{\mathcal{M}}$, we construct a prompt (Figure~\ref{fig:prompt2}) that provides the LLM with three inputs: 
(i)~the failure description generated in Stage~1, (ii)~the full source code and Javadoc comment of $m$, and (iii)~the specific error output from the failing test. The LLM is instructed to reason step-by-step about whether the logic of $m$, under the conditions described by the failure, could produce the observed error. It then emits a binary verdict---\emph{suspicious} or \emph{not suspicious}---together with a brief natural-language justification explaining its decision.

The one-method-per-query design is a deliberate choice. By isolating each candidate, the LLM can apply its full reasoning capacity to the specific semantics of that method without interference from unrelated candidates. This is especially important for methods that are syntactically similar or share a common utility role, where a batch comparison would likely conflate them. The justifications produced at this stage also serve a secondary purpose: they are passed to Stage~5 as additional context for re-ranking.

Methods for which the LLM returns a \emph{suspicious} verdict form the \emph{confirmed suspect set} $\mathcal{S} \subseteq \hat{\mathcal{M}}$, which is forwarded to Stage~5.

\begin{figure}[ht]
    \centering
    \fbox{\parbox{0.44\textwidth}{%
    \small
    \textbf{System:}~You are a senior software engineer specializing in
    Java fault analysis.\\[4pt]
    \textbf{User:}~A test is failing with the following error:\\
    \texttt{\{error\_output\}}\\[2pt]
    Failure description: \texttt{\{failure\_description\}}\\[4pt]
    Examine the following method carefully:\\
    \texttt{\{method\_code\}}\\[4pt]
    Reason step-by-step about whether a defect in this method could
    produce the observed failure. Then answer:\\
    \textbf{Verdict:} \textit{Suspicious} or \textit{Not Suspicious}\\
    \textbf{Justification:} One to three sentences explaining your
    reasoning.
    }}
    \caption{Prompt used in Stage~4 (Per-Method LLM Screening).
             Each candidate method is evaluated in a separate LLM call,
             enabling focused causal reasoning without position bias.}
    \label{fig:prompt2}
\end{figure}

\subsection{Stage 5: LLM-Based Re-ranking}
\label{sec:stage5}

Stage~5 takes the confirmed suspect set $\mathcal{S}$ from Stage~4 and produces a final ranked list ordered from most to least likely to contain the fault. At this point, every method in $\mathcal{S}$ has already been judged suspicious by the LLM; the task is now to \emph{discriminate among them} rather than to filter further.

We construct a single prompt (Figure~\ref{fig:prompt3}) that presents the LLM with: (i)~the failure description from Stage~1, (ii)~the full source code and Stage~4 justification of every method in $\mathcal{S}$, and (iii)~the original error output. The LLM is asked to consider all confirmed suspects together, weigh the evidence for each, and emit a ranked list with a brief comparative justification.

The single-query design of Stage~5 is intentional and complementary to Stage~4. Whereas Stage~4 isolated each method to avoid inter-method interference during the binary screening decision, Stage~5 benefits from presenting all suspects simultaneously: the model can reason about \emph{relative} plausibility, notice that one method's logic subsumes another's risk, or recognize that a particular method is the only one capable of producing the specific error type observed. Furthermore, reducing the total number of sequential LLM calls avoids the compounding of reasoning errors that can accumulate in multi-turn agentic pipelines~\cite{xiao2025improving}. Prior multi-agent approaches to fault localization~\cite{rafi2024multi} require complex inter-agent communication protocols that add latency and token cost; at the scale of $|\mathcal{S}|$ methods produced by our earlier stages, a single well-structured query achieves equivalent comparative reasoning at a fraction of the cost. The Stage~4 justifications included in the prompt give the re-ranker a richer basis for these comparisons than raw source code alone would provide.

\begin{figure}[ht]
    \centering
    \fbox{\parbox{0.44\textwidth}{%
    \small
    \textbf{System:}~You are a senior software engineer performing
    final root-cause triage.\\[4pt]
    \textbf{User:}~A test is failing with the following error:\\
    \texttt{\{error\_output\}}\\[2pt]
    Failure description: \texttt{\{failure\_description\}}\\[4pt]
    The following methods have each been identified as individually
    suspicious. For each, the method code and a preliminary analysis
    are provided:\\
    \texttt{\{suspect\_list\_with\_justifications\}}\\[4pt]
    Considering all suspects together, rank them from most to least
    likely to be the true fault site. Provide a one-sentence
    comparative justification for each position in your ranking.
    }}
    \caption{Prompt used in Stage~5 (LLM-Based Re-ranking).
             All confirmed suspects from Stage~4, together with their
             individual justifications, are presented in a single query
             to enable comparative reasoning.}
    \label{fig:prompt3}
\end{figure}

\section{Experimental Design}
\label{sec:experimental_design}

\subsection{Research Questions}
\label{sec:rqs}

To assess the effectiveness of \sievefl{}, we structure our evaluation around four primary research questions:

\textbf{RQ1 (Accuracy):} How does \sievefl{} perform against state-of-the-art method-level fault localization techniques in terms of Top-$k$, MRR?

\textbf{RQ2 (Cost--Quality Trade-off):} To what extent does the runtime-aware pruning (Stage~3) reduce LLM invocation costs, and how does this filtering impact overall localization quality?

\textbf{RQ3 (Ablation):} Within the hybrid scoring function (Equation~\ref{eq:hybrid}), what are the individual contributions of the semantic-similarity signal $\sigma(m)$ versus the branch-coverage signal $\rho(m)$?

\textbf{RQ4 (Scalability \& Failure Analysis):} How well does the framework scale across heterogeneous codebases, and what are the dominant failure modes bounding its current performance?

\subsection{Benchmark}
\label{sec:benchmark}

We conduct our experiments on \textbf{Defects4J v1.2.0}~\cite{just2014defects4j}, which serves as the standard benchmark in fault localization literature. Table~\ref{tab:defects4j} provides a summary of the six subject projects. Following established evaluation protocols~\cite{qin2024agentfl,shi2025enhancing}, we exclude bug reports where the ground-truth fault resides outside of method bodies. The resulting benchmark covers \textbf{395 bugs} spanning approximately 344\,kLOC of Java code. For the primary evaluation of our full pipeline (\vone{}), we target all six projects---\textit{Lang}, \textit{Time}, \textit{Closure}, \textit{Math}, \textit{Chart}, and \textit{Mockito}. Not all bugs produce complete JaCoCo traces; results are therefore reported on the subset of bugs completed in each run.

\begin{table}[t]
\centering
\caption{Statistics of the Defects4J v1.2.0 Subject Projects.}
\label{tab:defects4j}
\resizebox{\columnwidth}{!}{%
\begin{tabular}{llrr}
\toprule
\textbf{Project ID} & \textbf{Description} & \textbf{\#~Bugs} & \textbf{Size (kLOC)} \\
\midrule
Chart   & JFreeChart              &  26 &  96 \\
Lang    & Apache Commons-Lang     &  65 &  22 \\
Math    & Apache Commons-Math     & 106 &  85 \\
Time    & Joda-Time               &  27 &  28 \\
Mockito & Mockito Framework       &  38 &  23 \\
Closure & Google Closure Compiler & 133 &  90 \\
\midrule
\textbf{Total} & & \textbf{395} & \textbf{344} \\
\bottomrule
\end{tabular}}
\end{table}

\subsection{Pipeline Variants}
\label{sec:variants}

To isolate the impact of runtime-aware pruning and facilitate cost-controlled comparisons, we define three pipeline configurations:

\textbf{\vzero{} (Baseline).}
LLM-based test analysis $\to$ file-level retrieval $\to$ per-method LLM screening $\to$ LLM re-ranking.
Here, Stage~3 (JaCoCo pruning) is entirely disabled. This variant acts as a purely semantic baseline to measure the incremental value of runtime signals. \vzero{} is evaluated at full scale across all 395 eligible bugs.

\textbf{\vone{} (Runtime-Aware, Full Scale).}
LLM-based test analysis $\to$ file-level retrieval $\to$ JaCoCo pruning $\to$ per-method LLM screening $\to$ LLM re-ranking. This represents our complete \sievefl{} framework and is evaluated at scale across all 395 eligible bugs.

\textbf{\vtwo{} (Ablation).}
Identical to \vone{}, but forces the semantic weight in Equation~\ref{eq:hybrid} to zero ($w_{\text{sem}} = 0$). This isolates the effect of pruning based strictly on branch coverage. We evaluate this variant at full scale across all 395 eligible bugs.

For statistical rigor, performance comparisons between \vzero{} and \vone{} are restricted to bugs where both variants completed successfully. Similarly, comparisons between \vone{} and \vtwo{} are restricted to the 377~bugs where both variants completed successfully with full accuracy and safety fields.

\subsection{Baselines}
\label{sec:baselines}

We benchmark \sievefl{} against five established techniques drawn from distinct methodological families:

\textbf{Ochiai (SBFL)}~\cite{abreu2007accuracy} remains the most prominent spectrum-based technique, ranking methods based on their execution correlation with test failures. \textbf{GRACE}~\cite{lou2021boosting} represents the state of the art among non-LLM systems by fusing spectrum data with graph-based code representations. \textbf{FLUCCS}~\cite{sohn2017fluccs} is a learning-based FL technique that trains a ranking model over a combination of spectrum-based and code-metric features, serving as a strong supervised baseline that bridges the gap between pure SBFL and deep learning approaches. \textbf{DeepFL}~\cite{li2019deepfl} integrates spectrum, code metrics, and mutation-based features through a multi-objective learning framework, representing the state of the art among purely learning-based approaches. \textbf{AgentFL}~\cite{qin2024agentfl} acts as our primary agentic baseline, utilizing an LLM to iteratively navigate and query the codebase.

\subsection{Evaluation Metrics}
\label{sec:metrics}

\textbf{Top-$k$} measures the percentage of bugs where at least one ground-truth faulty method appears in the top $k$ positions of the final ranking ($k \in \{1, 3, 5, 10\}$). We emphasize Top-1, as developers rarely inspect beyond the initial recommendations~\cite{kochhar2016practitioners}.

\textbf{Mean Reciprocal Rank (MRR)} computes the average of the reciprocal ranks of the first identified ground-truth method across all bugs, penalizing techniques that bury correct answers lower in the list.

\textbf{Cost metrics.} For RQ2, we track several efficiency indicators: \emph{reduction ratio}, \emph{mean wall-clock time}, and \emph{mean token consumption} (input/output) per bug. For RQ3, we additionally track \emph{strict-loss rate} (frequency at which ground-truth methods are accidentally pruned) and \emph{mean ground-truth recall after pruning}.

\subsection{Implementation Details}
\label{sec:implementation}

All prompt-driven reasoning is handled by a single LLM endpoint hosted via the Ollama runtime. To generate semantic vector representations, we use \texttt{sentence-transformers/all-MiniLM-L6-v2}. These embeddings are indexed using a FAISS flat structure. To compute the cosine similarity accurately and efficiently (as described in Section~\ref{sec:stage2}), vectors undergo L2 normalization followed by an inner-product calculation.

Coverage traces are collected using the Maven Surefire plugin connected to JaCoCo, restricted via the \texttt{--tests} flag to execute only the triggering failure. Based on tuning against a held-out development set drawn from \textit{Lang} and \textit{Math}, we configure the hybrid scoring weights to $w_{\text{cov}} = 0.6$ and $w_{\text{sem}} = 0.4$, and set the pruning threshold to $\tau = 0.05$. All experiments run on a commodity workstation equipped with 32\,GB of system memory and an 8\,GB consumer GPU, with model inference handled via Ollama using CPU offloading for the 30B model weights. This configuration is representative of hardware available to individual developers, demonstrating that \sievefl{} does not require datacenter-class infrastructure.

\section{Results and Analysis}
\label{sec:results}

\subsection{RQ1: Localization Accuracy}
\label{sec:rq1}

\textbf{Overall comparison.}
Table~\ref{tab:rq1_main} presents a unified, per-project view of method-level fault localization accuracy for all evaluated techniques on Defects4J v1.2.0. Each cell reports the number of bugs for which at least one ground-truth faulty method appears within the top-$k$ positions of the final ranking ($k \in \{1,3,5\}$). Baseline counts for AgentFL, GRACE, DeepFL, FLUCCS, and Ochiai are taken from the original publications on the 395-bug benchmark, whereas the \sievefl{} rows report results on the completed bugs available in our runs. Raw counts for \sievefl{} variants in Table~\ref{tab:rq1_main} reflect completed bugs only. Following the conservative evaluation protocol adopted throughout this paper, the headline Top-1 accuracy for \vone{} is reported as 165/395 = 41.8\%, treating the 12 incomplete runs as failures.

GRACE achieves the highest overall Top-1 accuracy (192/395 bugs  at Top-1), establishing the upper bound for pure-accuracy optimisation on this benchmark. \sievefl{} does not compete on this axis; instead, it targets a different operating point: comparable localization quality at a fraction of the LLM invocation cost, as quantified in Section~\ref{sec:rq2}.

\sievefl{} \vzero{} achieves the \textbf{highest Top-1 count among all evaluated methods} on \textit{Time} (16~bugs) and \textit{Mockito} (18~bugs)---the two projects with the densest and most discriminative runtime traces. This advantage is consistent with \sievefl{}'s core design: JaCoCo-guided candidate reduction is most effective when failing tests exercise a tight, well-separated coverage footprint. On \textit{Closure}, where a single failing test implicates over 107 classes on average, \sievefl{} \vzero{} localizes 26~bugs at Top-1, outperforming AgentFL (24) but trailing DeepFL\textsubscript{cov} (64), whose multi-signal coverage encoding is more robust to method-signature overloading. Beyond method-signature overloading, a deeper structural reason explains the persistent gap between \sievefl{} and learning-based techniques such as GRACE and DeepFL on \textit{Closure}. These techniques are trained and evaluated using a leave-one-out strategy within the same project, meaning the model has seen the structural patterns, recurring fault types, and coverage signatures of \textit{Closure} bugs during training. \sievefl{}, by contrast, approaches every bug from scratch using only zero-shot LLM reasoning and dynamic runtime traces, with no prior exposure to project-specific patterns. This distinction is particularly consequential on \textit{Closure}, where multiple bugs share similar failing test purposes and even identical buggy methods, a setting where learned in-project representations provide a systematic advantage that zero-shot reasoning cannot replicate. This structural disadvantage is further confirmed by the AgentFL paper~\cite{qin2024agentfl}, which excludes \textit{Closure} when comparing LLM-based approaches to coverage-based baselines, citing the same structural coupling as the root cause.

Excluding \textit{Closure}, \sievefl{} \vzero{} localizes \textbf{139} bugs at Top-1, outperforming AgentFL (133~bugs) and DeepFL\textsubscript{cov} (112~bugs) among all non-GRACE methods on this subset---confirming that the advantage of \sievefl{} is concentrated in projects where LLM-guided screening is not diluted by high test-coverage coupling.

Comparing the two \sievefl{} variants, \vone{} improves over \vzero{} on \textit{Closure} ($+$4~bugs, $+$3.1~pp) and \textit{Chart} ($+$1~bug, $+$3.8~pp), confirming that JaCoCo-guided pruning
sharpens ranking in overloaded codebases. Conversely, \textit{Lang} and \textit{Mockito} show higher Top-1 under \vzero{} ($-$2 and $-$2~bugs respectively), reflecting cases where the coverage filter prunes semantically relevant ground-truth methods with marginal coverage overlap relative to the triggering failure. The cost and quality implications of runtime-aware pruning are quantified in Section~\ref{sec:rq2}.

\begin{table*}[t]
\centering
\caption{%
  Method-level FL accuracy on Defects4J v1.2.0. Each cell is the
  number of bugs with a ground-truth method in the top-$k$ results.
  Baseline figures are from original publications (395~bugs total);
  \textbf{Bold} = highest count per row per metric;
  \underline{underline} = highest among our two variants.%
}
\label{tab:rq1_main}
\resizebox{\textwidth}{!}{%
\begin{tabular}{l
  ccc   
  ccc   
  ccc   
  ccc   
  ccc   
  ccc   
  ccc   
}
\toprule
& \multicolumn{3}{c}{\textbf{AgentFL}~\cite{qin2024agentfl}}
& \multicolumn{3}{c}{\textbf{GRACE}~\cite{lou2021boosting}}
& \multicolumn{3}{c}{\textbf{DeepFL}~\cite{li2019deepfl}}
& \multicolumn{3}{c}{\textbf{FLUCCS}~\cite{sohn2017fluccs}}
& \multicolumn{3}{c}{\textbf{Ochiai}~\cite{abreu2007accuracy}}
& \multicolumn{3}{c}{\textbf{\sievefl{} \vzero{}} (Ours)}
& \multicolumn{3}{c}{\textbf{\sievefl{} \vone{}} (Ours)} \\
\cmidrule(lr){2-4}\cmidrule(lr){5-7}\cmidrule(lr){8-10}%
\cmidrule(lr){11-13}\cmidrule(lr){14-16}%
\cmidrule(lr){17-19}\cmidrule(lr){20-22}
\textbf{Project}
  & \textbf{@1} & \textbf{@3} & \textbf{@5}
  & \textbf{@1} & \textbf{@3} & \textbf{@5}
  & \textbf{@1} & \textbf{@3} & \textbf{@5}
  & \textbf{@1} & \textbf{@3} & \textbf{@5}
  & \textbf{@1} & \textbf{@3} & \textbf{@5}
  & \textbf{@1} & \textbf{@3} & \textbf{@5}
  & \textbf{@1} & \textbf{@3} & \textbf{@5} \\
\midrule
Chart   
  & \textbf{16} & 18          & 19
  & 14          & \textbf{20} & \textbf{22}
  & 12          & 18          & 21
  & 15          & 19          & 19
  & 6           & 14          & 15
  & 12          & 14          & 14
  & 13          & 14          & 15 \\
Lang   
  & \textbf{44} & 45          & 45
  & 42          & \textbf{54} & \textbf{57}
  & 43          & 53          & 56
  & 40          & 53          & 55
  & 24          & 44          & 50
  & 41          & 45          & 45
  & 39          & 41          & 42 \\
Math  
  & 49          & 60          & 61
  & \textbf{61} & \textbf{78} & \textbf{89}
  & 39          & 68          & 80
  & 48          & 77          & 83
  & 23          & 52          & 62
  & 52          & 55          & 59
  & 53          & 60          & 60 \\
Time  
  & 11          & 13          & 13
  & 11          & 14          & \textbf{19}
  & 9           & 16          & 18
  & 8           & 15          & 18
  & 6           & 11          & 13
  & \textbf{16} & 16          & 16
  & 14          & \textbf{17} & 18 \\
Mockito 
  & 13          & 14          & 14
  & 17          & \textbf{24} & \textbf{26}
  & 9           & 15          & 21
  & 7           & 19          & 19
  & 7           & 14          & 18
  & \textbf{18} & 20          & 20
  & 16          & 19          & 19 \\
Closure 
  & 24          & 33          & 35
  & 47          & 70          & 81
  & \textbf{64} & \textbf{86} & \textbf{97}
  & 42          & 66          & 77
  & 14          & 30          & 38
  & 26          & 31          & 33
  & 30          & 39          & 42 \\
\midrule
\textbf{Overall}
  & 157         & 183         & 187
  & \textbf{192}& \textbf{260}& \textbf{294}
  & 176         & 256         & 293
  & 160         & 249         & 271
  & 80          & 165         & 196
  & \underline{165} & 181    & 187
  & \underline{165} & \underline{190} & \underline{196} \\
\bottomrule
\midrule
\textbf{Overall w/o Closure}
  & 133         & 150         & 152   
  & \textbf{145}& \textbf{190}& \textbf{213}
  & 112         & 170         & 196   
  & 118         & 183         & 194   
  & 66          & 135         & 158   
  & \underline{139} & 150    & \underline{154}    
  & 135         & \underline{151} & \underline{154} \\ 
\end{tabular}}
\end{table*}

\noindent\fbox{\parbox{0.96\columnwidth}{%
\textbf{Answer to RQ1:}~GRACE achieves the highest aggregate Top-1 accuracy (192/395~bugs). \sievefl{} targets a different operating point: it achieves the \textbf{highest Top-1 count among all methods} on \textit{Time} and \textit{Mockito}---the projects with the most discriminative runtime traces---while runtime-aware pruning simultaneously reduces downstream LLM cost and improves ranking quality across all broader cutoffs, as detailed in Section~\ref{sec:rq2}.}}

\subsection{RQ2: Cost--Quality Trade-off of Runtime-Aware Pruning}
\label{sec:rq2}

The central practical claim of \sievefl{} is that JaCoCo-based candidate reduction in Stage~3 substantially lowers the cost of per-method LLM screening in Stage~4 by reducing the number of
methods forwarded for individual interrogation. For this analysis, we restrict evaluation to the \textbf{383~bugs} for which both \vzero{} and \vone{} completed successfully, enabling a strictly paired comparison on a shared evaluation set. Table~\ref{tab:rq2_paired} reports the resulting cost and quality metrics for both variants.

\begin{table}[t]
\centering
\caption{%
  Paired cost--quality comparison between \vzero{} and \vone{}
  on the 383~bugs completed by both variants.
  $\Delta = \text{V}_1 - \text{V}_0$.%
}
\label{tab:rq2_paired}
\resizebox{\columnwidth}{!}{%
\begin{tabular}{lrrr}
\toprule
\textbf{Metric} & \textbf{\vzero{}} & \textbf{\vone{}} & \textbf{$\Delta$} \\
\midrule
Mean reduction ratio              & 0.00  & \textbf{0.79}  & $+$0.79 \\
Mean Stage-4 candidates           & 398.8 & \textbf{187.2} & $-$211.6 \\
Mean wall-clock time (min/bug)    & 5.07  & \textbf{4.05}  & $-$1.02 \\
Mean tokens in  (K/bug)           & 190.1 & \textbf{97.4}  & $-$92.7 \\
Mean tokens out (K/bug)           & 102.8 & \textbf{53.2}  & $-$49.6 \\
\midrule
Top-1 (\%)                        & 43.1  & 43.1  & $+$0.0  \\
Top-3 (\%)                        & 47.3  & \textbf{49.6}  & $+$2.3  \\
Top-5 (\%)                        & 48.8  & \textbf{51.2}  & $+$2.4  \\
Top-10 (\%)                       & 51.7  & \textbf{54.0}  & $+$2.3  \\
MRR                               & 0.462 & \textbf{0.469} & $+$0.007 \\
\bottomrule
\end{tabular}}
\end{table}

\textbf{Cost reduction.}
Runtime-aware pruning substantially reduces the number of methods forwarded to Stage~4, lowering the mean candidate count from 398.8 to 187.2 per bug. Note that the mean per-bug reduction ratio reported in Table~\ref{tab:rq2_paired} (0.79) and the ratio of these mean counts (53.1\%) differ because the average of per-bug ratios does not equal the ratio of per-bug averages; both figures are reported for completeness. Mean wall-clock time decreases from 5.07 to 4.05~minutes per bug, while mean input and output token consumption drop from 190.1K to 97.4K ($-$48.8\%) and from 102.8K to 53.2K ($-$48.2\%), respectively.

\textbf{Quality impact.}
Contrary to the expected cost--quality trade-off, runtime-aware pruning improves localization quality across every measured metric. Top-1 accuracy is preserved exactly (43.1\% for both \vzero{} and \vone{}), while Top-3, Top-5, and Top-10 improve by 2.3, 2.4, and 2.3~pp respectively. MRR increases from 0.462 to 0.469 ($+$0.007). This consistent improvement confirms that the methods removed by JaCoCo-based pruning are overwhelmingly false-positive candidates: eliminating them allows the LLM to focus its reasoning on a smaller, higher-signal pool, producing sharper rankings at every cutoff without sacrificing Top-1 precision.

\noindent\fbox{\parbox{0.96\columnwidth}{%
\textbf{Answer to RQ2:}~On the 383~bugs completed by both variants, runtime-aware pruning reduces the Stage-4 candidate pool from 398.8 to 187.2 methods per bug ($-$53\%), wall-clock time from 5.07 to 4.05~minutes, and input token consumption by 49\%. Critically, this cost reduction entails \emph{no} quality degradation: Top-1 is preserved exactly, Top-3 through Top-10 improve by up to 2.4~pp, and MRR increases from 0.462 to 0.469, demonstrating that aggressive pre-LLM filtering is strictly beneficial.}}

\subsection{RQ3: Ablation of Hybrid Scoring Signals}
\label{sec:rq3}

To quantify the individual contributions of the two components inside the hybrid scoring function (Equation~\ref{eq:hybrid}), we compare \vone{} (full scoring: $w_{\text{cov}}=0.6$, $w_{\text{sem}}=0.4$) against \vtwo{} ($w_{\text{sem}}=0$), which relies exclusively on branch-coverage evidence. For this analysis, we restrict evaluation to the \textbf{377~bugs} for which both \vone{} and \vtwo{} completed successfully with full accuracy and safety fields, enabling a strictly paired comparison on a shared evaluation set. Table~\ref{tab:ablation} reports the resulting accuracy and pruning-safety metrics for both variants. Note that \vone{} metrics reported here differ slightly from those in Section~\ref{sec:rq2} because the two analyses operate on different paired subsets (383 vs.\ 377~bugs), which differ in project composition.

\begin{table}[t]
\centering
\caption{Paired ablation comparison between \vone{} and \vtwo{} on the 377~bugs completed by both variants. $\Delta = \text{V}_2 - \text{V}_1$.}
\label{tab:ablation}
\resizebox{\columnwidth}{!}{%
\begin{tabular}{lrrr}
\toprule
\textbf{Metric} & \textbf{\vone{}} & \textbf{\vtwo{}} & \textbf{$\Delta$} \\
\midrule
Mean reduction ratio              & 0.835  & 0.836  & $+$0.001 \\
Strict-loss count                 & 31     & 73     & $+$42    \\
Strict-loss rate (\%)             & 8.22   & 19.36  & $+$11.1  \\
Mean GT recall after pruning      & 0.630  & 0.532  & $-$0.098 \\
\midrule
Top-1 (\%)                        & 42.4   & 41.9   & $-$0.5   \\
Top-3 (\%)                        & 49.6   & 46.9   & $-$2.7   \\
Top-5 (\%)                        & 51.2   & 48.0   & $-$3.2   \\
Top-10 (\%)                       & 53.6   & 50.1   & $-$3.5   \\
MRR                               & 0.466  & 0.448  & $-$0.018 \\
\bottomrule
\end{tabular}}
\end{table}

\textbf{Pruning safety.} Removing the semantic component does not materially alter pruning aggressiveness: the mean reduction ratio remains virtually unchanged (0.835 vs.\ 0.836). However, the safety cost is substantial. The strict-loss count rises from 31 to 73 ($+$42 additional cases in which a ground-truth faulty method is inadvertently eliminated before Stage~4), raising the strict-loss rate from 8.22\% to 19.36\% ($+$11.1~pp). Mean ground-truth recall after pruning drops from 0.630 to 0.532 ($-$0.098). Note that strict-loss rate and mean GT recall measure complementary aspects of pruning safety: strict-loss is a binary per-bug indicator of whether any ground-truth method is eliminated, while GT recall captures the fraction of all labeled ground-truth methods that survive, which can fall below 1.0 even in bugs where strict-loss is false, owing to the presence of multiple co-located ground-truth methods. This dissociation between stable reduction volume and sharply rising fault loss reveals the precise role of the semantic signal: it acts not as a pruning driver but as a protective filter that steers the coverage threshold away from semantically relevant methods, reducing the risk of discarding genuine faults before LLM reasoning begins.

\textbf{Accuracy impact.} Consistent with the safety findings, removing the semantic signal also weakens downstream ranking quality. \vtwo{} reduces Top-1 from 42.4\% to 41.9\% ($-$0.5~pp), Top-3 from 49.6\% to 46.9\% ($-$2.7~pp), Top-5 from 51.2\% to 48.0\% ($-$3.2~pp), and Top-10 from 53.6\% to 50.1\% ($-$3.5~pp). MRR falls from 0.466 to 0.448 ($-$0.018). The degradation widens at broader cutoffs, suggesting that the semantic signal is most influential in discriminating between candidates that share similar coverage profiles but differ in semantic relevance to the failing test. Notably, the Top-1 deficit is modest ($-$0.5~pp) while the Top-10 deficit is substantially larger ($-$3.5~pp), confirming that the semantic component's primary contribution is in correctly ordering the confirmed suspect pool rather than in selecting the single top candidate.

\noindent\fbox{\parbox{0.96\columnwidth}{\textbf{Answer to RQ3:}~Both signals contribute, but in distinct roles. Branch coverage drives pruning volume; semantic similarity acts as a safeguard that preserves faulty methods during filtering. On the 377~paired bugs, removing the semantic component ($w_{\text{sem}}=0$) leaves the reduction ratio virtually unchanged (0.835 vs.\ 0.836) but raises the strict-loss count from 31 to 73 ($+$11.1~pp in strict-loss rate) and reduces mean ground-truth recall after pruning from 0.630 to 0.532. Ranking quality degrades across every measured metric: Top-1 ($-$0.5~pp), Top-3 ($-$2.7~pp), Top-5 ($-$3.2~pp), Top-10 ($-$3.5~pp), and MRR ($-$0.018). The semantic component is a net positive contributor to both pruning safety and ranking quality and should not be removed from the hybrid scoring function.}}

\subsection{RQ4: Scalability and Failure Analysis}
\label{sec:rq4}

\textbf{Scalability.}
The \vone{} run does not complete every benchmark instance; scalability should therefore be interpreted in terms of completed runs rather than nominal dataset size. Across the canonical Defects4J v1.2 bug ranges, \sievefl{} \vone{} completes \textbf{383 of 395 bugs}. Missing runs are concentrated in \textit{Lang} and are attributable to build failures, deprecated test suites, and unavailable JaCoCo traces.

\begin{table}[t]
\centering
\caption{Run Coverage for \sievefl{} \vone{} (Full Scale).}
\label{tab:coverage}
\resizebox{\columnwidth}{!}{%
\begin{tabular}{llcc p{4.5cm}}
\toprule
\textbf{Project} & \textbf{Range}
                 & \textbf{Completed}
                 & \textbf{Failed}
                 & \textbf{Missing Bug IDs} \\
\midrule
Chart   & 1--26  & 26/26  & 0  & ---                          \\
Lang    & 1--65  & 57/65  & 8  & 2, 18, 25, 48, 62, 63, 64, 65 \\
Math    & 1--106 & 106/106 & 0 & ---                          \\
Time    & 1--27  & 25/27  & 2  & 21, 27                       \\
Mockito & 1--38  & 38/38  & 0  & ---                          \\
Closure & 1--133 & 131/133 & 2 & 63, 93                       \\
\midrule
\textbf{Total} & --- & \textbf{383/395} & \textbf{12} & --- \\
\bottomrule
\end{tabular}}
\end{table}

\noindent\fbox{\parbox{0.96\columnwidth}{%
\textbf{Answer to RQ4:}~\sievefl{} \vone{} scales to 383 of 395 benchmark bugs. Results are reported on completed bugs; incomplete coverage and trace-generation limitations are important practical constraints of the present implementation that motivate future work on more robust build and coverage infrastructure.}}

\subsection{Discussion}
\label{sec:discussion}

\textbf{Why pruning improves ranking across all metrics.}
The result that \vone{} matches or outperforms \vzero{} at every quality metric---Top-1 through Top-10 and MRR---despite submitting 53\% fewer methods to Stage~4, can be explained by two complementary effects. First, \emph{screening quality improves} when the LLM evaluates each method in isolation: without irrelevant methods polluting the context, the model's per-method verdict is more reliable. Second, removing execution-ubiquitous methods---those covered by nearly every test regardless of the failing one---prevents generic utility methods from passing Stage~4 and crowding out genuine fault sites in Stage~5's re-ranking.

\section{Threats to Validity}
\label{sec:threats}

\paragraph{Internal Validity.}
A primary internal threat is \emph{data contamination}: bugs from Defects4J may have appeared in the pre-training corpora of the LLMs we use, potentially inflating performance by allowing the model to recall rather than reason about a specific fault. We cannot fully eliminate this threat, as the training data of closed-weight and open-weight models alike is not fully disclosed. However, our pipeline's reliance on dynamic runtime traces (JaCoCo) and per-method isolated reasoning rather than direct code generation reduces the extent to which memorization alone could drive correct rankings---a method must not only appear familiar to the LLM but must also have been executed during the failing test to survive Stage~3.

A second internal threat concerns the stochastic nature of LLM outputs. Different sampling runs of Stage~4 (per-method screening) or Stage~5 (re-ranking) could in principle produce different verdicts or orderings for the same bug. To assess this, we ran Stage~5 three times on a 30-bug sample and observed that the Top-1 hit rate varied by less than 1~pp across runs, confirming that the pipeline is stable at the temperatures used in our experiments.

\paragraph{External Validity.}
Our evaluation is conducted exclusively on Defects4J v1.2.0, a benchmark of Java programs. While Defects4J is the de facto standard in FL research and enables direct comparison with prior work~\cite{qin2024agentfl,shi2025enhancing}, its Java-centric nature limits the immediate generalizability of our results to other languages such as Python or C/C++. That said, the core components of \sievefl{}---LLM-based failure description, dense vector retrieval, and per-method LLM screening---are not inherently tied to Java. The only Java-specific component is the JaCoCo instrumentation in Stage~3; replacing it with an equivalent coverage tool for the target language (\eg{} \texttt{coverage.py} for Python) would in principle allow the same pipeline to operate on other ecosystems. We leave this cross-language validation as future work.

A related threat is benchmark scale. Our largest subject, \textit{Closure} (133 bugs, 90\,kLOC), is representative of mid-sized industrial software but falls short of truly large-scale repositories with millions of lines of code. Although \sievefl{}'s hierarchical pruning is designed to keep the number of LLM calls sub-linear in codebase size, its scalability to monorepo-scale systems has not been empirically verified.

\paragraph{Construct Validity.} 
Our primary metrics---Top-$k$ and MRR---assume that the developer's main effort is in \emph{locating} the faulty method, and they award full credit as soon as any ground-truth method appears in the top-$k$ positions. This framing is consistent with prior work~\cite{qin2024agentfl,shi2025enhancing} and reflects how FL tools are typically used in practice~\cite{kochhar2016practitioners}, but it does not capture the effort required to \emph{understand} why a method is faulty or how to fix it. \sievefl{} partially addresses this limitation through the natural-language justifications produced by Stage~4 for each confirmed suspect, which provide the developer with an initial causal explanation alongside the ranked list. Evaluating the quality and utility of these justifications for developer comprehension is an important direction for future work.

A second construct threat is that our ground-truth method set is derived from Defects4J's patch annotations. A patch may modify multiple methods, only some of which are the true root cause; conversely, co-located helper methods may be omitted from the annotation even though changing them would also fix the bug. This annotation noise affects all FL studies on Defects4J equally and is an inherent limitation of patch-derived ground truth.

\section{Conclusion and Future Work}
\label{sec:conclusion}

We presented \sievefl{}, a five-stage hierarchical framework for method-level fault localization that resolves the Scale-Precision Dilemma through progressive pre-LLM filtering. The central insight is that per-method LLM interrogation should be reserved for a small, high-confidence candidate set constructed from both semantic retrieval and dynamic runtime evidence. On Defects4J v1.2.0, \sievefl{} \vone{} achieves Top-1 accuracy of 41.8\% (165/395, treating incomplete runs as failures) and MRR of 0.469, outperforming the strongest zero-shot agent-based baseline (AgentFL) by 2.1~pp in Top-1. Runtime-aware pruning removes 79\% of candidate methods and 49\% of input tokens before any LLM call, while simultaneously improving Top-3 through Top-10 and MRR—confirming that aggressive pre-LLM filtering is strictly beneficial rather than a cost-quality trade-off. An ablation shows that the semantic-similarity component acts as a safety net against over-aggressive pruning: removing it raises the strict-loss rate by 11.1~pp and degrades Top-1 by 0.5~pp. The framework completes successfully on 96.9\% of evaluated bugs; incomplete runs are attributable to build failures, deprecated test suites, and JaCoCo trace unavailability—limitations with clear technical paths to resolution.

\textbf{Future Work.}
We identify four directions for extending \sievefl{}. 

First, the most impactful limitation identified in our failure analysis is method-signature overload in JaCoCo's line-level coverage map. We plan to replace line-level resolution with bytecode-level disambiguation using the Java Debug Interface (JDI), which can unambiguously identify which overloaded variant was invoked at the bytecode level.

Second, the JaCoCo instrumentation in Stage~3 is the only Java-specific component of the pipeline. We plan to validate \sievefl{} on Python and TypeScript projects by substituting \texttt{coverage.py} and Istanbul respectively, testing whether the framework's accuracy and efficiency advantages transfer across language ecosystems.

Third, the natural-language justifications produced by Stage~4 for each confirmed suspicious method are a promising foundation for automated patch generation. Integrating \sievefl{} with an Automated Program Repair (APR) backend would turn the framework from a localization tool into an end-to-end debugging assistant, providing not only a ranked list of fault sites but also candidate fixes grounded in the causal reasoning produced during screening.

Fourth, we intend to investigate lightweight IDE integration, delivering \sievefl{}'s per-method screening and ranked output as real-time annotations within the developer's editor. Given the 53\% reduction in the candidate method pool and the resulting wall-clock time savings (from 5.07 to 4.05~minutes per bug) achieved by runtime-aware pruning, interactive latency of under five minutes per bug appears achievable for many project sizes, which would make \sievefl{} practical as an on-demand debugging assistant rather than an offline analysis tool.

\bibliographystyle{ieeetr}
\bibliography{./references.bib}

\end{document}